\documentclass{procmult}
\usepackage{graphicx}

\usepackage{color}
\usepackage{ulem}

\begin{document}
\title{{\huge Gravitational-wave extraction from neutron-star oscillations} }
\author{ %
  {\large 
    S. Bernuzzi$^{\dag, *}$,
    %
    %
    L. Baiotti$^\ddag$,
    G. Corvino${}^{\dag,+}$,
    R. De Pietri${}^\dag$,
    A. Nagar$^{\S}$ 
  }
}

\institute{ %
  {\normalsize %
    $^\dag$ Universit\`a di Parma and 
    INFN, Gruppo Collegato di Parma,
    43100 Parma, Italy 
    \\ 
    $^*$ Theoretical Physics Institute, University of Jena, 07743 Jena, Germany 
    \\ 
    %
    %
    $^\ddag$ Yukawa Institute for Theoretical Physics,
    Kyoto University, Kyoto 606-8502, Japan
    \\ 
    $^+$ Max-Planck-Institut f\"ur Gravitationsphysik,
    Albert-Einstein-Institut, Potsdam-Golm, Germany
    \\ 
    $^\S$ Institut des Hautes Etudes Scientifiques, 91440 Bures-sur-Yvette, France,
    and INFN, Sezione di Torino, Torino, Italy and ICRANet, 65122 Pescara, Italy 
    \\
  }
}


\maketitle
\abstract{
  We compare different gravitational-wave extraction methods used in
  three-dimensional nonlinear simulations against linear 
  simulations of perturbations of spherical spacetimes with matter.
  We present results from fully general-relativistic simulations of a system 
  composed by an oscillating and non-rotating star emitting gravitational 
  radiation. Results about the onset of non-linear effects are also 
  shown. 
}

\section{Simulation setup and wave extraction}

%
%
We considered a stable equilibrium star model perturbed 
with an axisymmetric quadrupolar fluid perturbation. 
Assuming the perfect fluid approximation for the matter,  
the star is described by a polytropic equation of state of the 
type\footnote{We use standard dimensionless units $c=G=M_\odot=1$
  unless otherwise specified, for clarity.}
$p=K \rho^{\Gamma}$ with $K=100$, $\Gamma=2$ and central rest-mass density $\rho_c=1.28\times
10^{-3}$. The gravitational mass and the radius are $M = 1.4 M_\odot$ and $R=9.57$. This model has
been widely used in the literature (see e.g. \cite{Font:2001ew}).  The perturbation consists in an
axisymmetric pressure perturbation: the angular pattern is proportional to the $(2,0)$ spherical
harmonic, while the radial profile has no nodes in the star interior.  Different perturbation
amplitudes are considered: $\lambda=\{0.001,0.01,0.05,0.1\}$, named $\lambda_0$,$\ldots$, 
$\lambda_3$. After imposing the fluid perturbation, the \emph{linearized} (Hamiltonian) constraint
is solved under the hypothesis of conformal flatness to obtain consistently the perturbed
metric. See \cite{Baiotti:2008nf,Bernuzzi:2008fu} for a detailed description and discussion.
 
%
%
The same initial data were then evolved with two different codes.
The {\sc PerBaCCo} code, described in \cite{Bernuzzi:2008fu}, was used to
solve in the time domain the (1+1) equations for the linear (even parity) 
perturbations of a spherical spacetime with matter.
The resolution of the (radial) 1D grid
is $\Delta r=0.032$, which corresponds to $300$ points inside the star.
Gravitational waves (GWs) are directly computed via the Zerilli-Moncrief function $\Psi^{\rm (e)}(t)$ (see below).
The {\tt Cactus-Carpet-CCATIE-Whisky} code was instead used
to solve the complete three-dimensional (3D) nonlinear hydrodynamics
equations coupled with Einstein's field equations within in the ``3+1'' formalism.
See \cite{Baiotti:2008ra} for a general description of the code and the recent results obtained.
The Cartesian 3D grid consists in 3 fixed refinement levels (boxes) with spacing (in each direction)
$\Delta x = 0.5$ or $\Delta x = 0.25$ or $\Delta x = 0.125$ for three resolutions used.

%
%
The extraction of gravitational radiation from a numerically computed spacetime is a 
non-trivial task .
The current state-of-the-art methods employed by the numerical-relativity community are the
\emph{metric} extraction {\it \'a la} Abrahams-Price \cite{Abrahams:1995gn} and the \emph{curvature}
Newman-Penrose extraction via the Weyl scalar $\psi_4$ \cite{Newman:1961qr}. If $h^{\ell m}$ are
the multipoles of the GW degrees of freedom, the metric extraction provides the Zerilli-Moncrief
$\Psi^{\rm (e)}_{\ell m}(t)$ and the Regge-Wheeler $\Psi^{\rm (o)}_{\ell m}(t)$ functions which are
directly related to $h^{\ell m}$:
\begin{equation} 
  \label{eq:h_from_psie}
  h^{\ell m}(t) = \frac{N_\ell}{r}
  \left(\Psi^{\rm (e)}_{\ell m}(t) + {\rm i} \Psi^{\rm (o)}_{\ell m}(t)\right) \ \ ,
\end{equation}
where $N_\ell=\sqrt{(\ell+2)(\ell+1)\ell(\ell-1)}$.  On the other hand if the curvature extraction
is used, the GW multipoles must be recovered from the Weyl scalar $\psi_4^{\ell m}(t)$ using the
relation
\begin{equation}
  \label{eq:h_from_psi4}
  \ddot{h}^{\ell m}(t) = \psi_4^{\ell m}(t) \ \ .
\end{equation}
Because of the finite size of the numerical grid, GWs wave are extracted at finite radius. 

In the following we analyze the performances of these methods in the ``simple'' system composed by a
single oscillating star. All the results are compared with the 1D code which gives more accurate
data for the waves. Our work, while differing from both, is motivated by the same ideas of
\cite{Pazos:2006kz,Shibata:2003aw}, namely testing the GW extraction methods with perturbative
results and using an oscillating neutron star as a test-bed.  It represents the first complete
analysis of such a kind employing simulations with the {\tt Whisky} code.

\section{1D waveforms: wave identikit}
\label{sec:pert}

\begin{figure}[t]
  \centering
  \includegraphics[width=\textwidth]{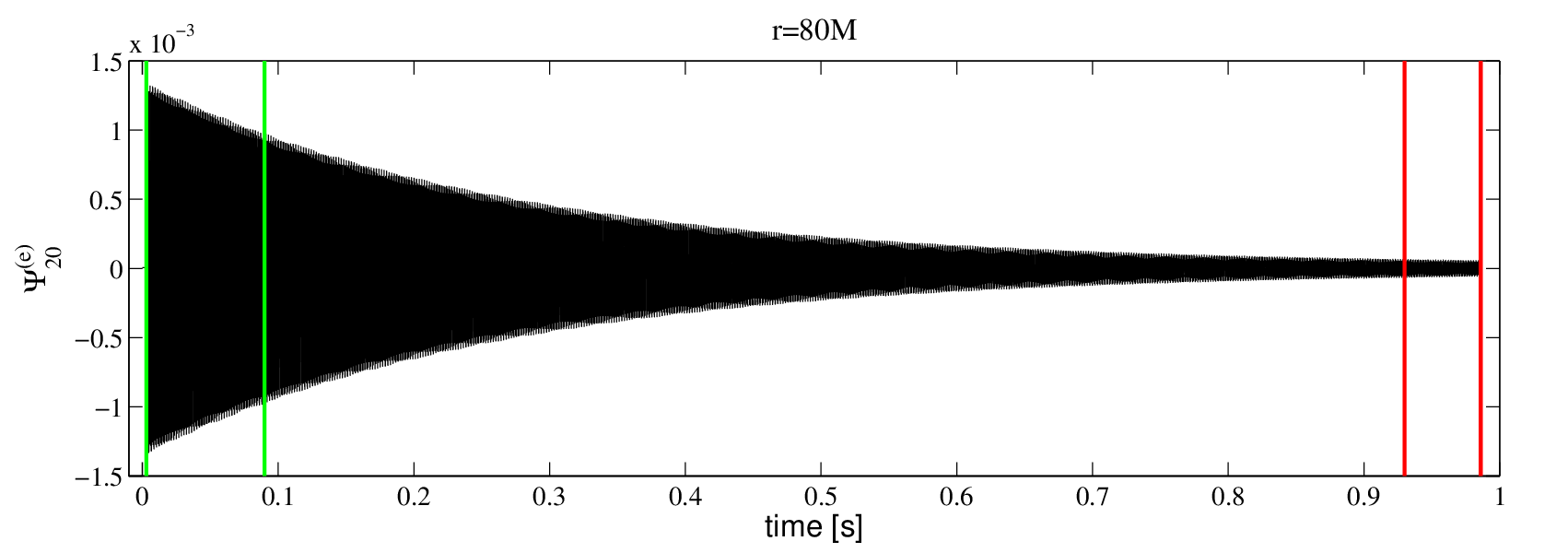}\\
  \includegraphics[width=0.49\textwidth]{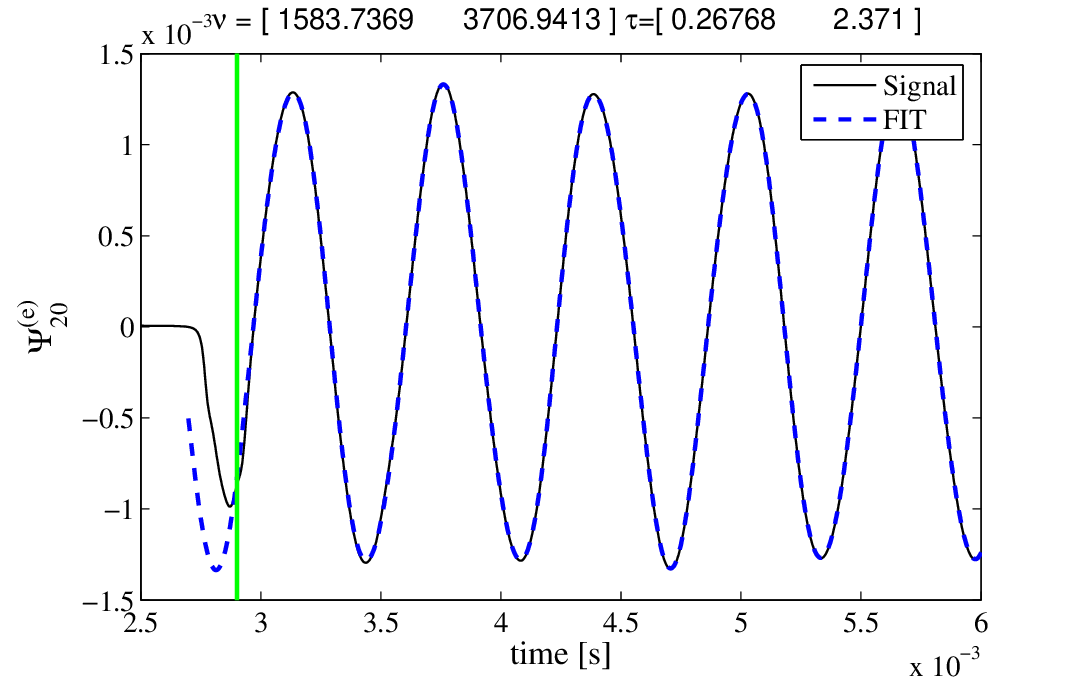}
  \includegraphics[width=0.49\textwidth]{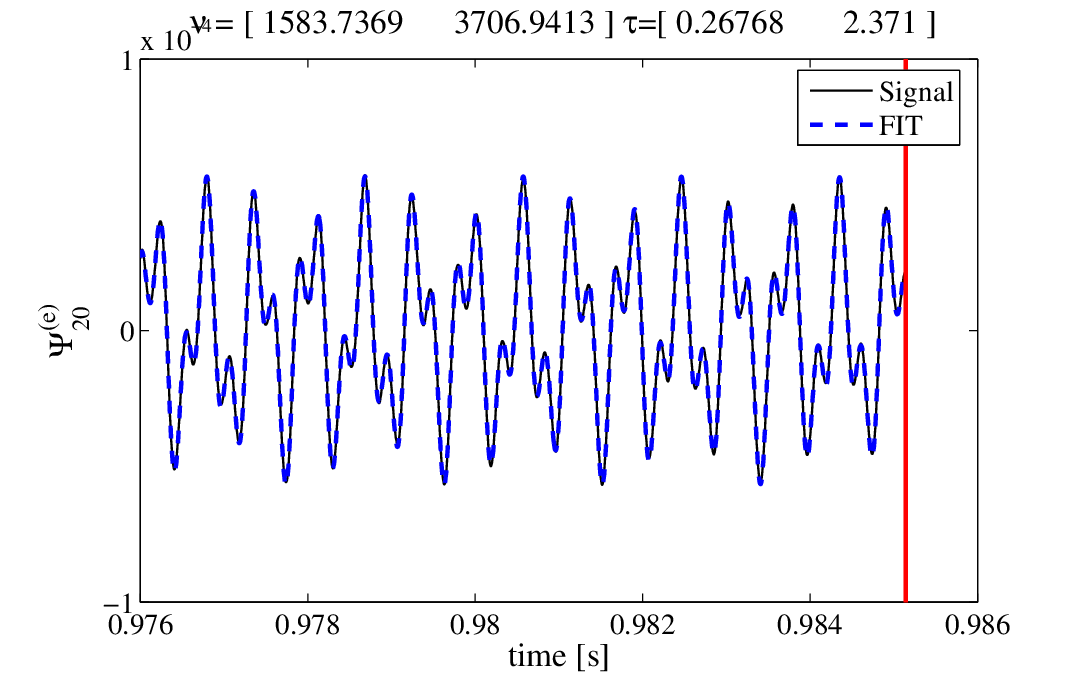}\\
  \includegraphics[width=0.49\textwidth]{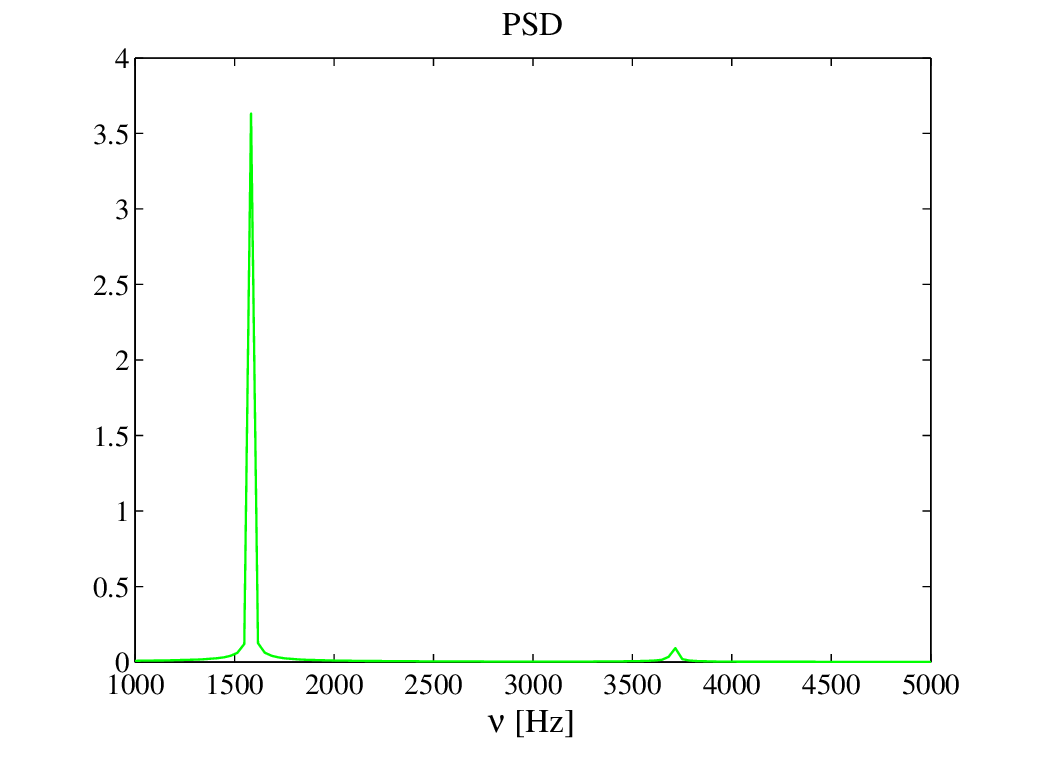}
  \includegraphics[width=0.49\textwidth]{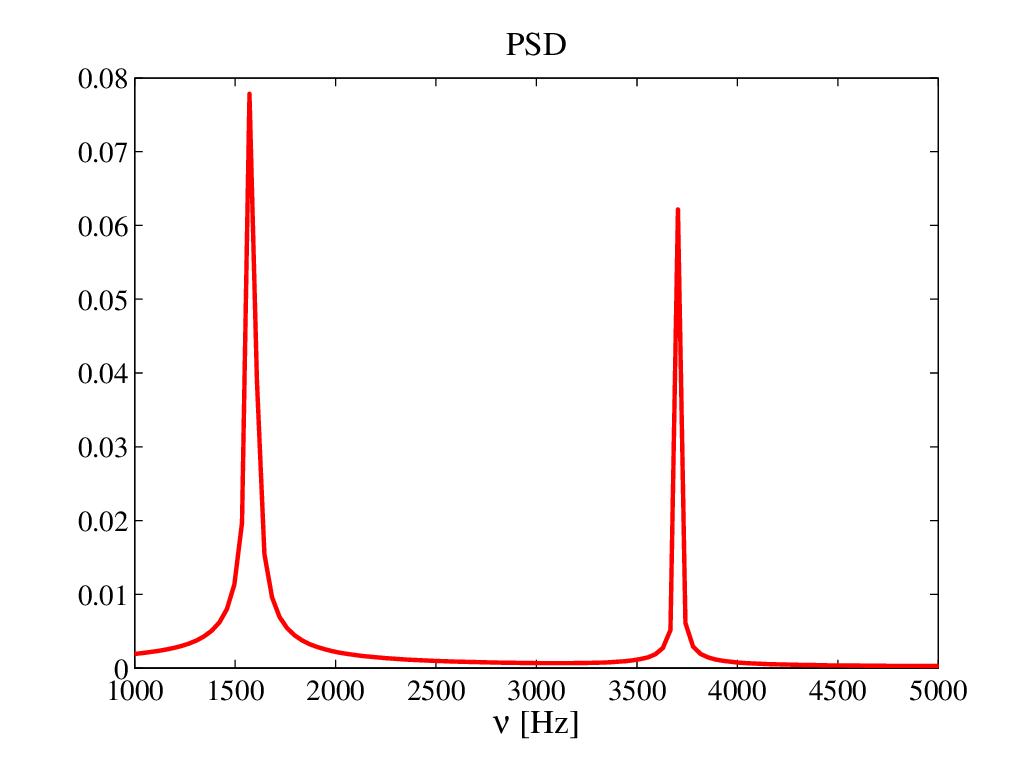}
  \caption{ {\small Analysis of the waves produced with the 1D perturbative code. 
    The top panel shows the entire waveforms $\Psi^{\rm (e)}_{20}(t)$. 
    In the middle panels a blow-up of the waveform (black solid line) and the fit 
    (blue dashed line)
    are shown at early (left) and late (right) time: 
    a single fit with the QNMs template in Eq.~\ref{eq:qnmfit} and two modes perfectly 
    reproduces the entire waveform.
    The bottom panels show the power-spectrum density taken, respectively, from the waveform's 
    time series in the interval $\Delta t \sim [0,0.9]$ s (left, in green) and 
    in the interval  $\Delta t \sim [0.9,1]$ s (right, in red): 
    the frequency of the $p_1$ mode is clearly present only at late time.} }
  \label{fig:pertwaves:anal}
\end{figure}

\begin{figure}[t]
 \centering
 \includegraphics[width=0.49\textwidth]{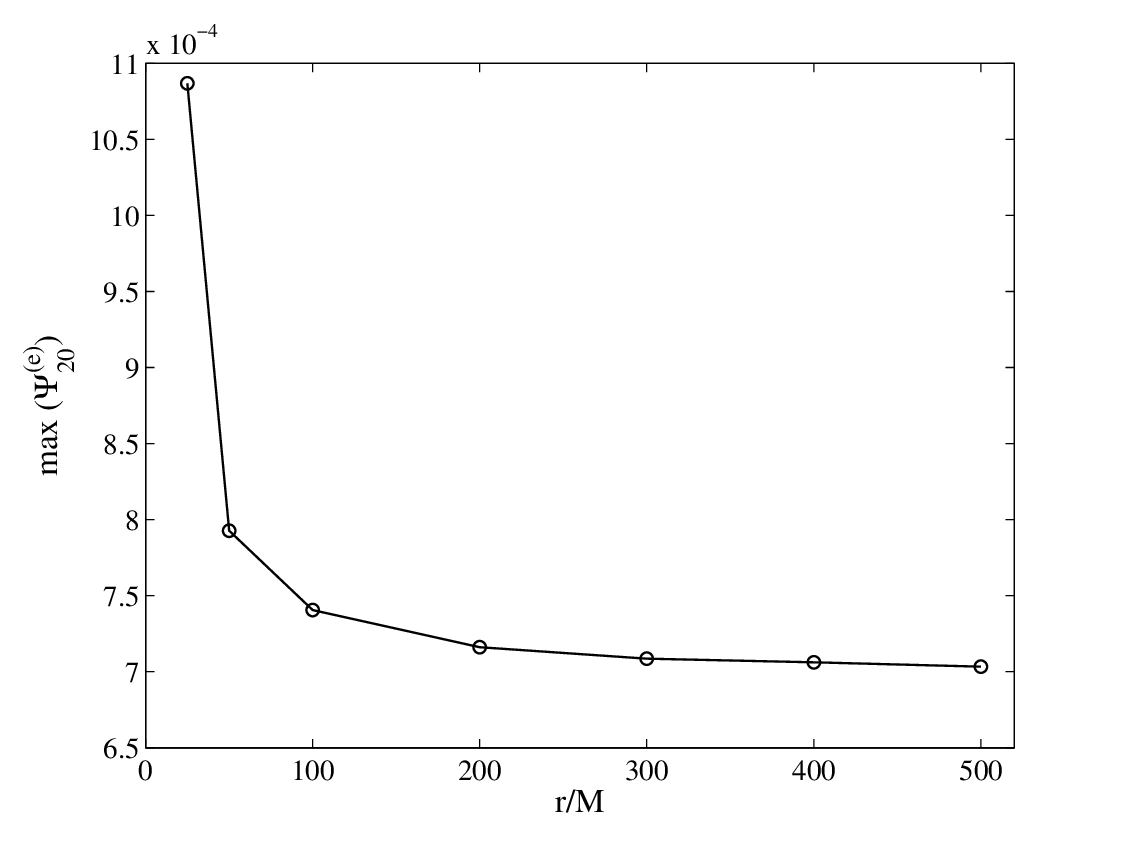}
 \includegraphics[width=0.49\textwidth]{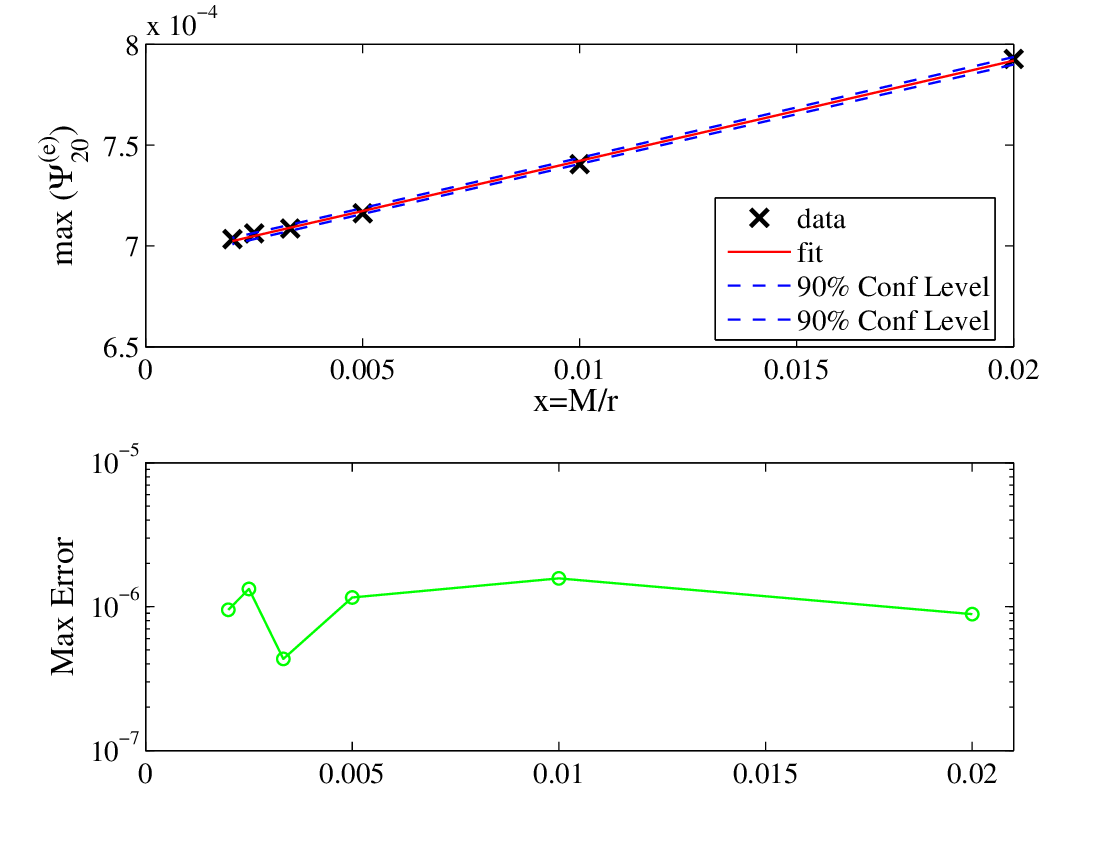}
 \caption{ {\small Analysis of finite-radius extraction effects on 1D perturbative waveforms. 
   The left panel shows the maximum of $\Psi^{\rm (e)}_{20}(t)$ extracted at different radii. 
   The right panels show the result of a linear fit with the template in Eq.~\ref{eq:fexfit}.} }
 \label{fig:pertwaves:fex}
\end{figure}

We give here a detailed ``identikit'' of the waves emitted 
by the system considered and computed using the 1D perturbative simulations.
This waves will serve as a starting point to compare with nonlinear
simulations described in the following sections.

The data we are presenting came from a very long simulation with final time 
of about $1$~s. The full waveform is plotted in the top panel of Fig.~\ref{fig:pertwaves:anal}.
It can be described by two different phases: (i) an initial transient,
of about half a GW cycle, say up to $t\simeq 3$~ms, related to the setup of
the initial data (see the black solid line in the left-middle panel), 
followed by (ii) a quasi-harmonic oscillatory phase, where the matter
dynamics are reflected into the stellar quasi-normal modes (QNMs):
\begin{equation}
  \label{eq:qnmfit}
  \Psi^{\rm (e)}_{20}(t) \simeq \sum_{n=0}^N A_{n} \cos\left(2\pi\nu_{n} t +\phi_{n}\right)\exp\left(-\alpha_{n} t\right) \ \ .
\end{equation}

From the Fourier spectrum of $\Psi^{\rm (e)}_{20}(t)$ over a time
interval from $3$ to about $30-90$ ms (see left-bottom panel),
we found that the signal is dominated
by the fundamental mode of oscillation of the star, the $f$-mode, at frequency $\nu_0=1581$ Hz, 
with a very weak contribution of the first pressure or $p$-mode (at frequency around $\nu_1\simeq 3724$ Hz).
The accuracy of these frequencies obtained from Fourier 
analysis on such long time series has been
checked in Ref.~\cite{Bernuzzi:2008fu} and is better than 1\%.
The frequency of the $f$-mode agrees with that of Ref.~\cite{Font:2001ew}
within $2$\%.
On the interval considered the waves can be perfectly  
represented by a one-mode expansion, i.e. Eq.~\ref{eq:qnmfit} with $N=1$.
As a result of a standard least-squares nonlinear fit with a one-mode template 
we obtained the frequency $\nu_{0}=1580.79\pm0.01$ Hz, which is perfectly
consistent with that obtained via Fourier analysis.
For the damping time, we estimated
$\alpha_{0}=3.984\pm0.066$ $\rm sec^{-1}$ and thus $\tau_{0}\equiv\alpha_{02}^{-1}\simeq0.25$ s.
The global residual ($l^\infty$ norm) of the fit is of the order of $10^{-6}$.

If we consider  the entire duration (1~s) of the signal it is clear that a one-mode expansion 
is not sufficient to reproduce accurately the waveform (see Figure~\ref{fig:pertwaves:anal}).
The Fourier analysis of the whole waveform reveals the presence of the $p_1$-mode, which has 
longer damping time, and the entire signal 
has been fitted with a two-mode QNM expansion.
The estimated values for the frequency of
the $f$-mode and the  $p_1$-mode are $\nu_0=1583.737\pm0.002$ and
$\nu_1=3706.941\pm0.001$, respectively, and the corresponding dumping times
$\tau_0=0.268$~s and $\tau_1=2.37$~s, with errors respectively of
the order of $0.1$\% and $2$\%. The frequencies are consistent with those computed
via Fourier analysis as well as via the dumping times. This result is particularly 
significant for a time-domain code.

Figure~\ref{fig:pertwaves:anal} shows the extraction result at (isotropic) radius 
of $r=80M$, 
which is the same extraction radius used for the comparison with the 3D results. However 
in the 1D simulations we can easily reach up to $r=500M$.
We checked (see Figure~\ref{fig:pertwaves:fex}) the convergence of the waves with the
extraction radius using as a reference point the maximum of
$\Psi^{(\rm e)}_{20}(t)$. This point can be accurately fitted, as a function
of the extraction radius, with
\begin{equation}
  \label{eq:fexfit}
  \max_{t}{\left(\Psi^{\rm (e)}_{20}(t,r)\right)}\simeq a^{\infty}+\frac{a^1}{r} \ .
\end{equation}
The extrapolated quantity $a^\infty$ allows an estimate of the error
related to the extraction at finite distance.
The difference in amplitude of a wave extracted at $r>200M$ with respect to
a wave extracted at infinity is less than $2\%$ and frequency seems not to be affected (within our accuracy) by
finite extraction effects.

\section{3D waveforms: comparison in the linear regime}
\label{sec:3Dlin}

\begin{figure}[t]
  \centering
  \includegraphics[width=0.49\textwidth]{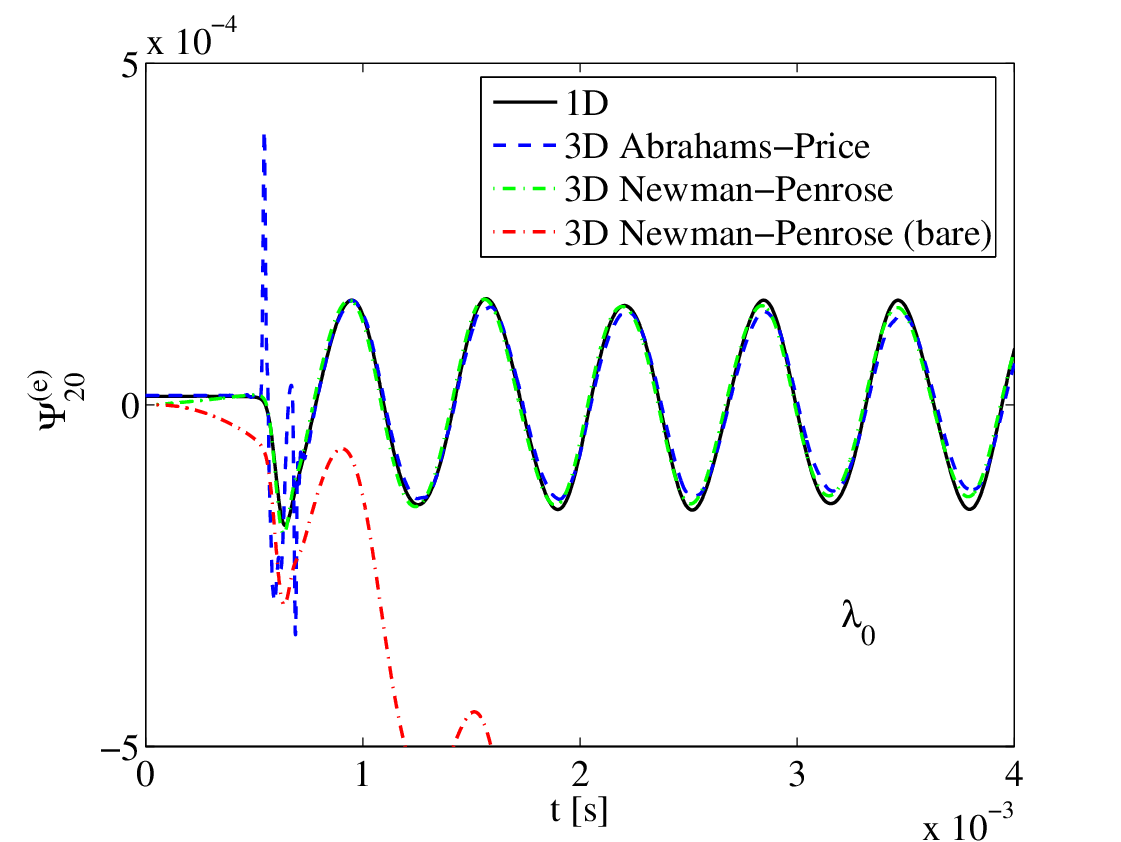}
  \includegraphics[width=0.49\textwidth]{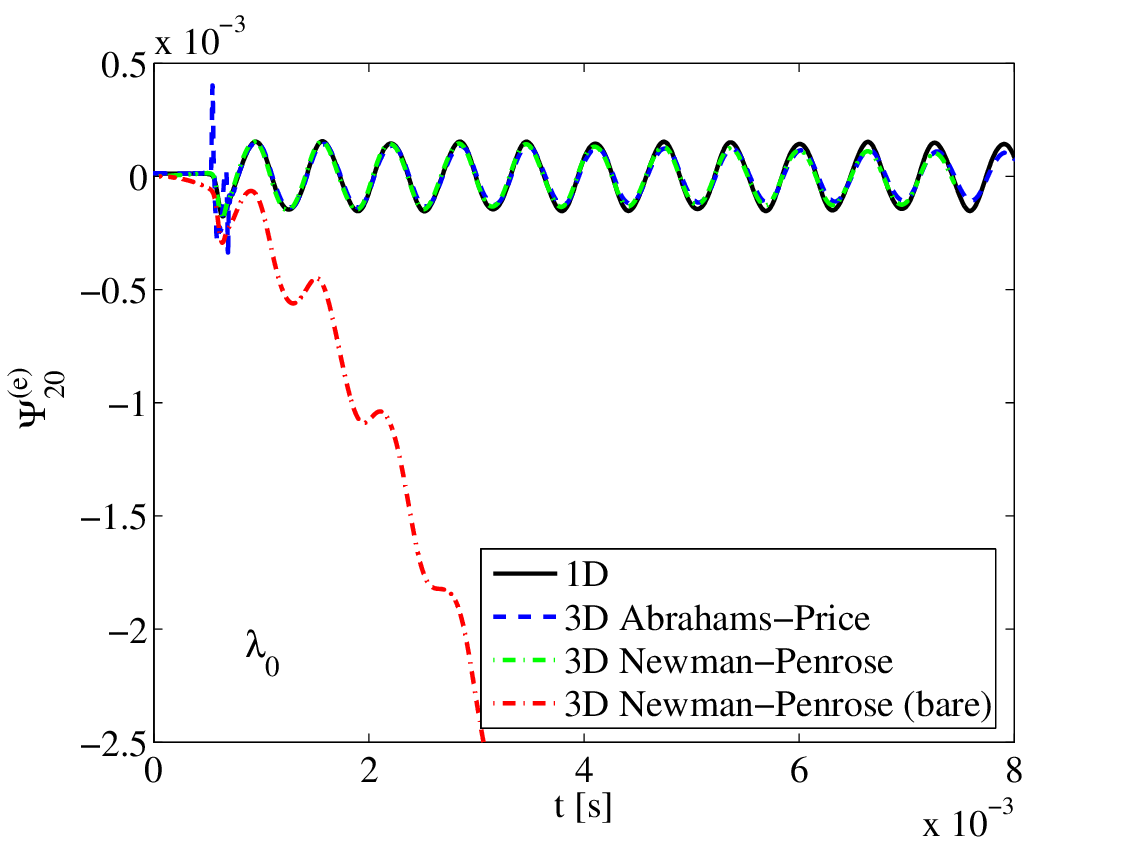}
 \caption{ {\small Comparison between perturbative and 3D waveforms extracted both 
   with the Abrahams-Price \emph{metric} method and Newman-Penrose \emph{curvature} 
   method. The results refer to the evolutions of the initial data 
   with the smallest perturbation amplitude
   $\lambda_0=0.001$. 
   The left panel shows $\Psi^{\rm (e)}_{20}(t)$ at early times, 
   while the right panel shows the entire time series. 
   See text for explanations.} }
 \label{fig:comp}
\end{figure}

We focus first on the data relative to the 
smallest initial perturbation amplitude $\lambda=\lambda_0=0.001$. The 3D simulations 
were performed for a simulation time of about 20 ms.
Figure~\ref{fig:comp} summaries the results of our findings. The black solid line is the 
waveform computed with the perturbative code and presented in the previous section.
We assume it to be the exact one.
The waveform extracted from the 3D simulation with the Abrahams-Price method is 
the blue dashed line in the figure. The only difference with the perturbative result
is the presence of a ``burst'' of radiation at the early times (see left panel) 
$0.5<t<1$ ms. In \cite{Baiotti:2008nf} we discussed in detail this feature of the metric wave
extraction. The main conclusions are the following.
{\it (i)} It is unphysical: it linearly grows with the extraction radius and produces a 
violation of the \emph{linearized} Hamiltonian constraint at the wave-extraction level.
{\it (ii)} It is related to an inaccurate implementation of the Zerilli-Moncrief 
function in the 3D code and
it can be reproduced in the perturbative code introducing ``by hand'' 
inaccuracies (very low resolution, low-order approximations for finite differences).
{\it (iii)} It is absent in vacuum simulations, 
e.g. binary black holes, where higher grid resolutions can be used.
Apart from the junk the metric waveform is perfectly superposed (until $t\sim 7$ ms)
to the exact one; frequency and amplitude are consistent within the error bars.

The GWs extracted via the $\psi_4$ scalar require the solution of Eq.~\ref{eq:h_from_psi4} 
to obtain the GW multipole (or equivalently the Zerilli-Moncrief function).
The procedure we followed is described in Ref.~\cite{Damour:2008te} and consists in 
computing a double integral of $\psi_4^{\ell m}$ (the ``bare'' metric waveform) and then 
correcting it by subtracting a polynomial floor.   
The red dashed-dotted line in Fig.~\ref{fig:comp} is the ``bare'' waveform: 
as suggested by the figure, and differently from Ref.~\cite{Damour:2008te}, 
we found that a quadratic polynomial is here necessary to align the waveform 
(see the green dashed-dotted line). Once this tuning has been done the $\psi_4$ waveform is 
exactly superposed to the perturbative one and, differently from the metric waveform, it is 
free from the {\it unphysical} burst of junk radiation. 

In summary we can conclude that, in the linear regime, both extraction methods give 
a very good agreement with the ``exact'' waveform but neither is free from 
drawbacks. On one hand the metric extraction contains a burst of junk radiation at early times
that can compromise the study of phenomena in the first 5 ms of simulation 
(e.g. the neutron-star collapse to black hole or neutron-star $w$-modes). On the other hand
the curvature extraction requires tuning of the integration constants.

\section{3D waveforms: the nonlinear regime}
\label{sec:3Dnonlin}

\begin{figure}[t]
  \centering
  \includegraphics[height=50mm, width=0.49\textwidth]{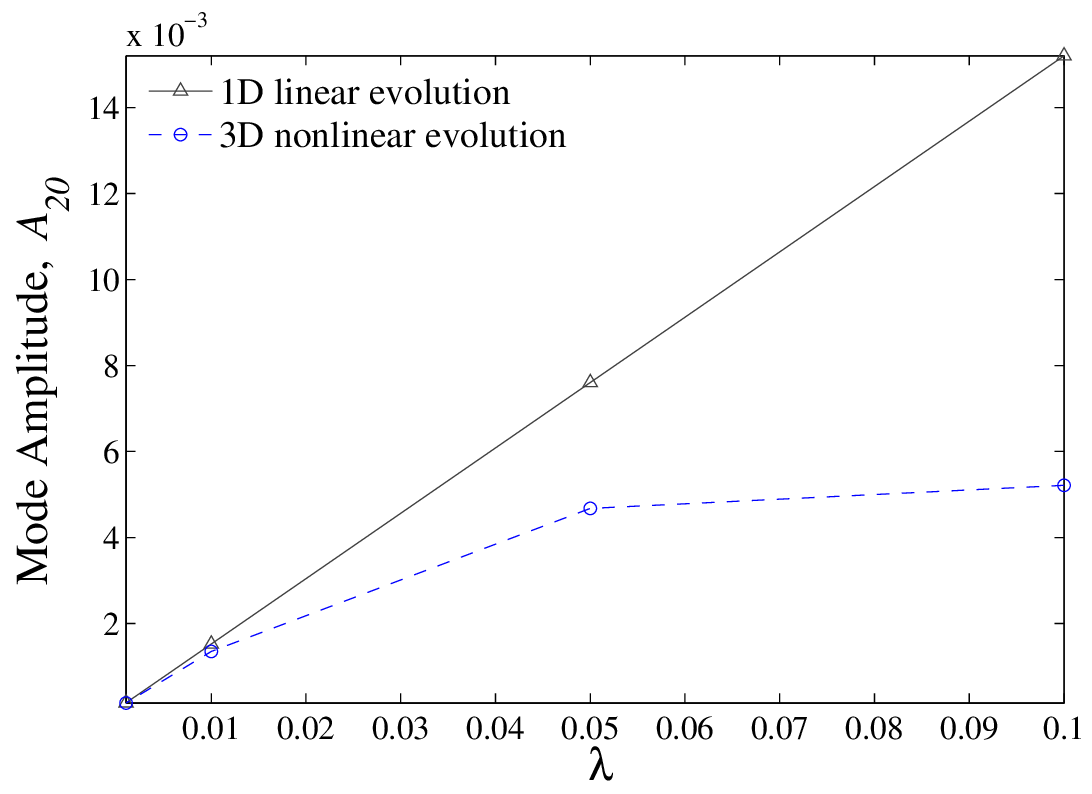}
  \includegraphics[height=50mm, width=0.49\textwidth]{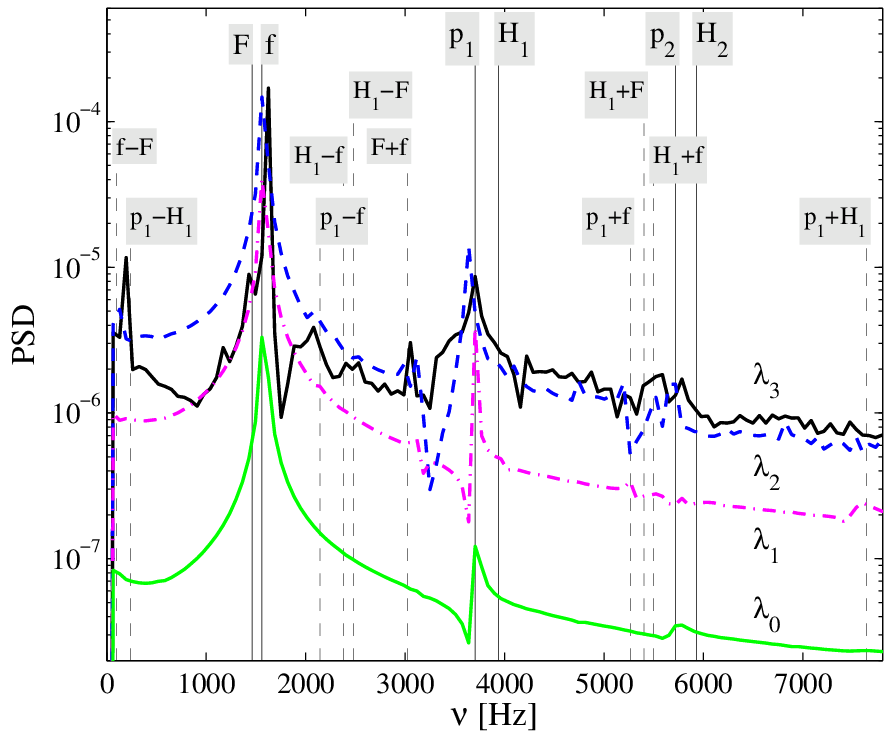}
  \caption{ {\small Nonlinear effects in the waves computed with the 3D code.
    The left panel shows the amplitude of the $(2,0)$ mode, 
    determined by fitting Eq.~\ref{eq:qnmfit} with $N=1$, 
    as a function of the initial perturbation amplitude $\lambda$:
    while the linear scaling of the linear code is perfect, 
    in the nonlinear case the deviations, 
    that can be seen already for small values of $\lambda=\lambda_1=0.01$, become
    conspicuous for $\lambda>\lambda_1$.
    The right panel shows the power-spectrum density of the integrals of the rest-mass density 
    projected trough the $(2,0)$ spherical harmonic. For larger values of $\lambda$
    more frequencies are present because of nonlinear mode couplings.} }
 \label{fig:nonlineff}
\end{figure}

Finally we discuss nonlinear effects in the 3D waveforms. Previous studies 
can be found for example in \cite{Passamonti:2007tm} (second-order perturbation theory) and 
in \cite{Dimmelmeier:2002bm} (CFC approximation of Einstein equations). We
confirm those results without using approximations.
Referring to the left panel of Fig.~\ref{fig:nonlineff}, we found that the amplitude of the
$(2,0)$ mode (computed fitting Eq.~\ref{eq:qnmfit}) 
scales linearly with the initial perturbation amplitude $\lambda$ - as expected -, but
it progressively decreases in the nonlinear case due to excitation of other channels. 
The nonlinear effects are already visible for $\lambda=\lambda_1$ (10\% difference 
in $A_{20}$) and quickly increase for $\lambda_2$ (38\%) and $\lambda_3$ (66\%). 
We note that the frequency of the wave, instead, is essentially not affected by nonlinearities 
with our setting.

Furthermore we considered the Fourier transform of the rest-mass density 
projected (multiplied and integrated on the 3D-space) on the spherical harmonic $(2,0)$. 
As the right panel of Fig.~\ref{fig:nonlineff} shows, for $\lambda=\lambda_0$ 
(green line) only the $\ell=2$ $f$ and $p_1$ (and $p_2$) mode frequencies are present, while
for higher values of $\lambda$ the spectra are rich of other frequencies.
Most of the frequencies can be associated to the radial ($\ell=0$) fundamental frequency 
$F$ (1466 Hz), its overtone $H_1$ (5955 Hz), and to \emph{combination tones} between the radial and the 
$\ell=2$ frequencies: $\nu_{\rm nnlin}=\nu_0 \pm \nu_2$.
This analysis prove that mode couplings are present in the dynamics of the system already for small initial perturbation amplitudes, but they affect strongly only the amplitude of the GWs.


\section*{Acknowledgments}

Computations have been performed on the INFN Beowulf clusters 
{\tt Albert} at the University of Parma and on the Peyote and
Damiana clusters at the Albert-Einstein-Institut. 



\end{document}